\UseRawInputEncoding
\documentclass[twocolumn,amsmath,amssymb,pre,superscriptaddress,longbibliography]{revtex4-2}

\usepackage{graphicx}
\usepackage{dcolumn}
\usepackage{bm}
\usepackage{float}

\usepackage[autostyle]{csquotes}

\usepackage{graphbox,graphicx}

\usepackage{lipsum}
\usepackage{multirow}
\usepackage{color}
\usepackage{xcolor}

\usepackage{hyperref}
\hypersetup{
    colorlinks=true,
    urlcolor=blue
}

\begin{document}

	\title{Jamming Crossovers in a Confined Driven Polymer in Solution}
	
	\author{Setarehalsadat Changizrezaei}
	\email{schangiz@uwo.ca}
	\affiliation{Department of Physics and Astronomy, The University of Western Ontario, London, Canada}
	\author{Mikko Karttunen}
	\email{mkarttu@uwo.ca}
 	\affiliation{Department of Physics and Astronomy, The University of Western Ontario, London, Canada}
	\affiliation{Department of Chemistry, The University of Western Ontario, London, Canada}
	\author{Colin Denniston}%
	\email{cdennist@uwo.ca}
	\affiliation{Department of Physics and Astronomy, The University of Western Ontario, London, Canada}

	
	\date{\today}

\begin{abstract}

We use lattice-Boltzmann molecular dynamics (LBMD) simulations to study the compression of a confined polymer immersed in a fluid and pushed by a large spherical colloid with a diameter comparable to the channel width. We examined the chain's deformation with both purely repulsive and weakly attractive Lennard-Jones (LJ) potentials applied between the monomers. The sphere's velocity was varied over 3 orders of magnitude. The chain is in a non-dense state at low pushing velocities for both repulsive and attractive monomer interactions. 
When the velocity of the spherical colloid exceeds a threshold $v^*$,
the back end of the chain transitions to a high density state with low mean square monomer displacement (MSD) values. The front end, however, remains in a non-dense state with high MSD indicating a pseudo two-state coexistence. This crossover is also revealed through volume per monomer and MSD as a function of the sphere's velocity. We also studied polymer dynamics by investigating folding events at different times.

\end{abstract}

\maketitle

\section{Introduction}

Polymers confined in nanochannels can
undergo dramatic conformational changes such as compression, twisting and bending~\cite{C7SM02413D,doi:10.1021/ma400674q, Mishra2017-pe,10.1063/1.1903923,CHEN2021124340,doi:10.1080/01411594.2010.534657}. 
Confined conditions are very common in biology and
play often a crucial role in phenomena observed, e.g., in DNA condensation, protein folding and dynamics,
and chromatin organization
\cite{Zimmerman1996-ko,Odijk1998-vc,5ref_p,7ref_p,8ref_p,Cino2012-ug,6ref_p}. In addition to importance in biological systems, conformational changes have also been used for applications such as DNA sequencing, size-dependent filtration, and single molecule manipulation and analyses in nanofluidic devices~\cite{2ref_p,3ref_p,4ref_p,Reisner2005-dt,Reisner2012-sx, Ollila2014-xg, doi:10.1021/acs.analchem.0c03868}. 

As the above indicates, it is important to investigate the statistical and dynamic properties of confined polymers under different conditions, and transient and steady-state non-equilibrium phenomena are of particular importance~\cite{9ref_p, 10ref_p,11ref_p,12ref_p,13ref_p,Chen2017-ez,Bleha2020-zv}.
Examples of out-of-equilibrium systems include 
the compression and extension dynamics of polymers and more complex systems such as DNA-like molecules interacting with non-DNA binding and DNA-binding proteins, where the DNA is confined into a cylindrical container and subjected to the action of a piston~\cite{1ref,Bleha2020-zv, Cifra2021-bs,Chen2021-ev}, and investigations of the kinetics of DNA collapse in a nanoslit~\cite{2ref}.

Based on a theory introduced by de Gennes and co-workers~\cite{doi:10.1051/jphys:0197700380108500,doi:10.1063/1.434540}, there are two equilibrium behaviours for a flexible self-avoiding walk (SAW) chain in a tube of diameter $D$, depending on the ratio of $R_\mathrm{F}/D$, where $R_\mathrm{F}$ is the Flory radius in a bulk solution. When $R_\mathrm{F} \ll D$, the polymer behaves essentially as it does in a bulk solution.  When 
$R_\mathrm{F} > D$
the chain can be thought of as a succession of impenetrable and hydrodynamically uncorrelated \enquote{blobs} of radius $D$. In this model, the length of the tube occupied by the chain scales as $R_{\parallel} = Na(a/D)^{2/3}$, where $a$ is the monomer size and $N$ is the 
degree of polymerization.
Inside each blob, excluded volume effects are maintained and the number of monomers per blob $(g)$ scales as $D = a g^\nu$, where $\nu$ is the Flory exponent. Further confinement leads to a third regime, referred to as Odijk confinement~\cite{doi:10.1021/ma00242a015}.

Important information about the elasticity of a macromolecule can be obtained from the restoring force profiles obtained during stretching experiments. Entropic elasticity of a polymer chain
is often described using two
models: the freely jointed chain (FJC)
model and the worm-like chain (WLC) model,
 see, e.g., the book of Doi and Edwards~\cite{Doi1986-cz} for details of the models.

In the FJC model, a polymer is represented as $N$ discrete Kuhn segments, each having length $l_\mathrm{k}$ (referred to as the Kuhn length). These segments are connected by flexible joints, and the model assumes the absence of any long-range interactions between them. 
The WLC model, 
a continuum formulation of the 1949 Kratky-Porod model~\cite{https://doi.org/10.1002/recl.19490681203}, describes a polymer chain as a continuous chain incorporating
bending stiffness, but neglecting any discrete structure. The model describes polymrs using a characteristic length scale, the persistence length $P$, arising from the bending stiffness. 
At length scales smaller than $P$  the polymer can be considered as a semiflexible chain, and
on scales exceeding $P$, the orientational order of the polymer segments decays exponentially. For semiflexible chains, 
the ratio $D/P$, where $D$ is the channel size,
can be used as an additional measure for the degree of confinement~\cite{de_Gennes,polym8080296}.  For $D/P \gg 1$, the polymer is in the de Gennes regime, which can be considered as a string of "blobs". On the other hand, for $D/P \ll 1$, the chain is in the Odijk regime.

Experimental studies of confined polymers under non-equilibrium conditions have been conducted using optical and magnetic tweezers, as well as atomic force microscopy~\cite{KRIEGEL201726,https://doi.org/10.1038/nmeth.1218}. For example, Khorshid {\it et al.}~\cite{18ref_p} investigated the dynamics of DNA molecules confined in a nanochannel and pushed by a sliding bead. They found that the DNA undergoes transient dynamics and reaches a well-defined highly compressed steady-state. They also observed that below a threshold speed $v^*$ the chain slid through the channel close to its equilibrium extension. At speeds above $v^*$, the chain was partially compressed close to the bead while the rest of the chain remained at its equilibrium concentration. At very high speeds, the entire chain became compressed. Further work on the effect of confinement on chain conformation can be found for, example, in Refs.~\cite{radhakrishnan2022compression, https://doi.org/10.1038/srep18438,PhysRevLett.95.268101}.

Confined driven systems have also been studied using computer simulations.
Hayase \textit{et al.}~\cite{PhysRevE.95.052502}  performed Langevin simulations
and showed that a sequence of recurring structural transitions occurs when
a chain in the Odijk regime ($D/P < 1$) is pushed into a channel.
The polymer transformed from a random
configuration into an ordered helical structure when it 
was being pushed into the channel. 
Upon further compression, the helix transformed into a  double-fold random deflection.

Bernier \textit{et al.}~\cite{doi:10.1021/acs.macromol.7b02748} studied the compression of a strongly confined semiflexible chain in a square channel with $D \ll P$.
They used Brownian dynamics (BD) simulations with a gasket translating at a fixed sliding speed impinging on the nanochannel extended chain
and found that repeated hairpin folds are formed as the chain is compressed by the moving gasket. In addition to being even deeper in the Odijk regime, Bernier \textit{et al.} also used a much lower $P$ in relation to the contour length of the polymer $L$, than Hayase (whose chains more closely resembled actin than DNA)~\cite{PhysRevE.95.052502}, thus exploring a significantly different confinement regime (very different $D/L$ as well as smaller $D/P$).

Zeng {\it et al.}~\cite{PhysRevE.105.064501}
investigated single nanochannel-confined semiflexible polymers compressed against a barrier in a homogeneous flow field. The barrier was approximated as perfectly porous to the fluid flow so that flow was assumed uniform up to the barrier. In this case, a polymer pushed against the stationary barrier, which is the end of the circular cross-section nanochannel,  by a steady flow field was argued to be equivalent to a polymer pushed at constant speed by a moving (piston) barrier in an immobile fluid. For stiff chains at low compression and regime $D < P$, the chain self-organizes into repeated hairpin folds. This was also observed by Bernier \textit{et al.}~\cite{doi:10.1021/acs.macromol.7b02748}.
However, coils (whole aggregate of loops, not just single circular loop) were formed in addition to the folds, that is,
coils coexisted alongside folds for stiff chains at higher flow, 
in contrast to the findings of Bernier \textit{et al.} Chain organization was retained even when the chain persistence length was comparable to the channel width. As Bernier \textit{et al.}~\cite{doi:10.1021/acs.macromol.7b02748} performed their simulations using nanochannels with square cross-section for similar values of stiffness and flow and did not find any coiled organization, it was concluded that configurations are geometry-dependent and it was suggested that having sharp corners causes an energy barrier that prevents coil formation.

Chen and Wei~\cite{CHEN2021124340} studied the deformations and dynamics of ring polymers with different confinements in compression and relaxation processes using  Langevin simulations.
They found that the compression and relaxation processes are dependent on the degree of confinement. A weakly confined flexible ring in the de Gennes regime was compressed into condensed blobs. This behaviour was also observed for linear polymers.  
In the relaxation process, the flexible ring relaxed back to its initial state. The strongly confined semiflexible ring in the Odijk regime ($D/P < 0.3$) undergoes a series of helix-collapse deformations, including parallel deflection, double helix, collapse, multiple-folded helix structures. When the piston was pulled back, the semiflexible ring extended from the compacted structure to a randomly folded structure.
For $1 > D/P > 0.3$, where the bending energy and excluded volume interactions are not strong enough to prohibit backfolding, the ring is buckling and packed into four-strands helix structures in the compression process. Considering the relaxation process,  it relaxes to four-stranded random structures, similar to the strong confined rings ($D/P < 0.3$). 

Roth\"orl \textit{et al.}~\cite{Rothorl2022-lm} used Langevin simulations to study
knot formation in a confined double-stranded DNA (dsDNA) while being pushed by a piston through a nanochannel.
As their major finding they reported non-monotonous dependence of knot formation on the persistence length. In addition, 
knot formation was dependent on the rate of the applied driving force (a piston). 

The current study focuses on conformational changes of a polymer confined in a nanochannel when it is being pushed by a spherical colloid with diameter comparable to the channel width.
We use hybrid LBMD simulations to capture the full hydrodynamics including thermal fluctuations~\cite{Ollila2011-wg}; hydrodynamic effects were not present in the works dicussed above. We also vary the speed of the sphere to study different polymer crossovers.

\section{Modeling}

\begin{figure}[t]
	\begin{center}
		\includegraphics[trim={330 0 0 0},scale=0.38]{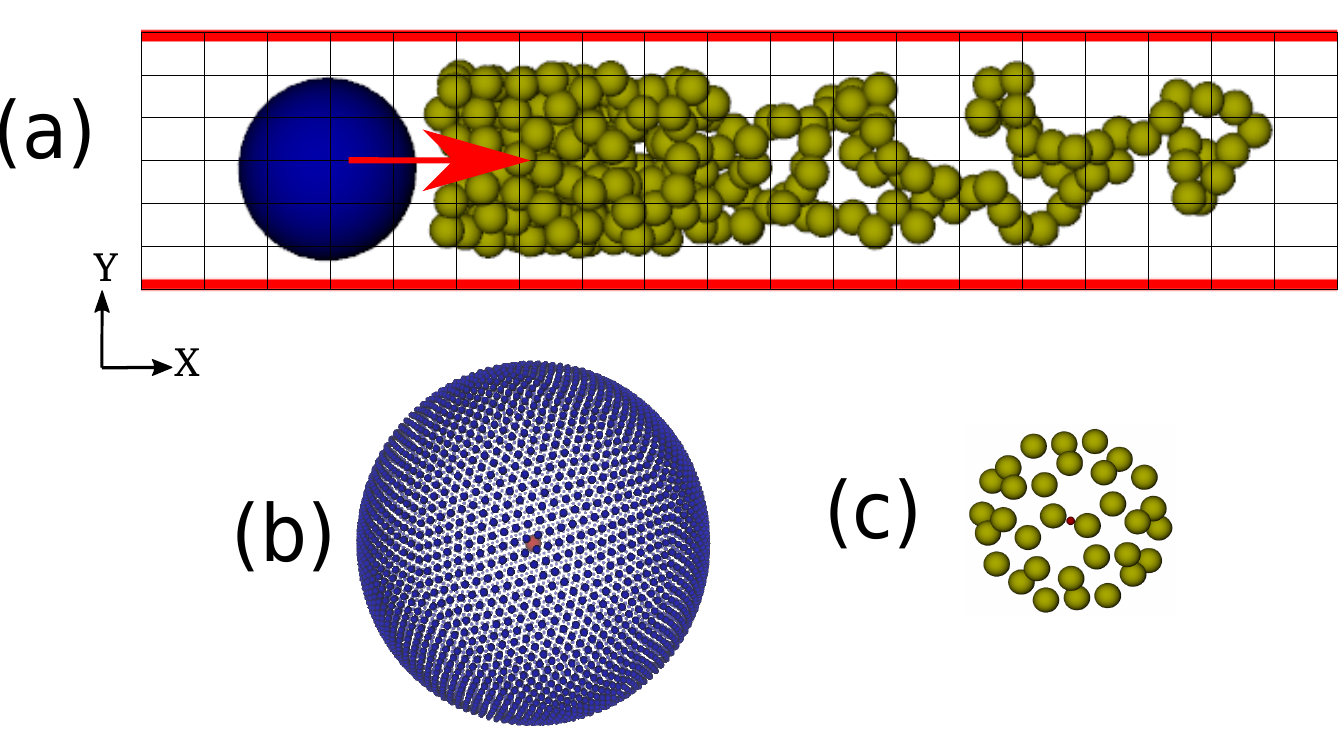}
		\caption{a) 
  Schematic of the $yx$-plane. Periodic boundary conditions are imposed in the $x$-direction, and fixed walls in the $y$- and $z$-directions, which are represented by the double red lines at the top and the bottom of the channel. The red arrow indicates the spherical colloid's direction of motion along the channel. The chain is in mustard and the colloid in blue. The polymer is in the de Gennes regime. b) The spherical colloid and c) monomer are comprised of spherical shell of 6,252 nodes and 60 nodes, respectively. The red dots are the MD atoms that are used for applying the LJ potential between the monomers and the spheres, and the FENE potential between the monomers.}
	\label{box}
	\end{center}
\end{figure}

In the LBMD approach, particles are modeled using MD
and the fluid with the lattice-Boltzmann (LB) method. 
Figure~\ref{box}(a) shows a schematic of our system.  
Fixed boundary conditions were used in the $y$-and $z$-directions and periodic boundary conditions in the $x$-direction. The channel cross-section (square) width was 25\,nm and the channel length 500\,nm (the channel is considerably longer than shown in Figure~\ref{box}).
All the simulations were performed using LAMMPS~\cite{Thompson2022-lp}. 

In the MD part, the bead-spring model was used for the coarse-grained polymer. 
The linear polymer chain consisted of 255 monomers.  Non-consecutive monomers along the chain interact via a Lennard-Jones (LJ) interactions
\begin{equation}
U_{LJ}(\sigma,\epsilon; r) = 4\epsilon \left[ \left(\frac{\sigma}{r} \right)^{12} - \left(\frac{\sigma}{r} \right)^6 + A
  \right] \Theta(r_c-r),
\end{equation}
where $r=|\vec{r}_i-\vec{r}_j|$ is the distance between non-adjacent monomers, $\sigma$ and $\epsilon$ are the LJ distance and energy scales, $r_c$ is the cutoff, $\Theta$ is the Heaviside step function, and $A$ is chosen so that $U_{LJ}(r_c)=0$.  We study two cases: (i) the purely repulsive LJ (the Weeks-Chandler-Andersen (WCA) potential~\cite{Weeks1971-ra}) where $r_c=2^\frac{1}{6}\sigma$ (and $A=1/4$) and; (ii) the weakly attractive case where $r_c=2.5 \sigma$.  In both cases, we used $\sigma\!=\!4.0\,\mathrm{nm}$, and $\epsilon = 1.0\,\mathrm{ag} (\mathrm{nm})^2 (\mathrm{ns})^{-2}$.  

Consecutive monomers along the chain were connected by 
the finitely extensible non-linear elastic (FENE)~\cite{46ref_p} bond potential.
The total potential energy is the sum of the FENE and LJ parts,
\begin{equation}
\label{FENE}
		U(r)= -\frac{1}{2} KR_0^2\log\left(1-\frac{r^2}{R_0^2}\right) + U_{LJ}(\sigma_F,\epsilon_F; r),
\end{equation}
where $r=|\vec{r}_i-\vec{r}_{i\pm 1}|$ is the distance between adjacent monomers, 
$K=100\, \mathrm{ag}\, \mathrm{nm}^{-1}$, and $R_0 = 10\, \mathrm{nm}$ is the maximum bond extension.  The $LJ$ part of the potential is purely repulsive and cutoff at $r_c=2^\frac{1}{6}\sigma_F$ (and $A=1/4$), $\sigma_F\!=\!4.0\,\mathrm{nm}$, and $\epsilon_F = 4.14195\,\mathrm{ag} (\mathrm{nm})^2 (\mathrm{ns})^{-2}$.

An additional WCA potential was applied between the monomers of the polymer and the pushing sphere to
prevent overlap. For that, we chose $\epsilon_\mathrm{ms} = 8.0\,\mathrm{ag} (\mathrm{nm})^2 (\mathrm{ns})^{-2}$ and $\sigma_\mathrm{ms} = 18\, \mathrm{nm}$.
These parameters are set large enough to ensure that no overlaps occur.

A repulsive LJ potential 
was also applied between the central atoms of the colloidal sphere/monomers (the red dots in Fig.~\ref{box}) and the fixed boundaries.
We chose $\epsilon_\mathrm{s-wall} = 4.0 \, \mathrm{ag} (\mathrm{nm})^2 (\mathrm{ns})^{-2}$ and $\sigma_\mathrm{s-wall}= 11 \,\mathrm{nm}$ for interaction between the sphere and the walls, and $\epsilon_\mathrm{m-wall} = 4 \,\mathrm{ag} (\mathrm{nm})^2 (\mathrm{ns})^{-2}$
and $\sigma_\mathrm{m-wall} = 4.0 \, \mathrm{nm}$ for interaction between the monomers and the walls.

To incorporate hydrodynamic interactions between the particles and the solvent, 
we used the $D_3Q_{15}$ LB model~\cite{29ref_p, 32ref_p, 33ref_p}, which is a 15-velocity model  on a cubic lattice to simulate fluid dynamics in three dimensions. 
The discretized version of the Boltzmann equation is then used to solve for the fluid motion.
For the simulations, the thermal LB method of Ollila \textit{et al.}~\cite{Ollila2011-wg} 
was used through the LB fluid package~\cite{27ref_p,Ollila2011-wg,DENNISTON2022108318} in LAMMPS~\cite{Thompson2022-lp}.
In the approach of Ollila \textit{et al.}, the lattice-fluid acts as a heat bath for the MD particles, 
that is, the polymer and the sphere are present in a nanochannel with full hydrodynamics and thermal fluctuations.
The mass and momentum conservation can be expressed as
\begin{equation}
	\begin{split}
		&\partial_t \rho + \partial_\alpha\left(\rho u_\alpha\right)=0  \\
		& \partial_t \rho + \partial_\beta(\rho u_\alpha u_\beta) = -\partial_\alpha P_{\alpha \beta} + F_\alpha \\
		&\!\! +\! \partial_\beta [\eta (\partial_\alpha u_\beta \!+\! \partial_\beta u_\alpha \!-\! \frac{2}{3} \partial_\gamma u_\gamma \delta_{\alpha \beta} ) \!+\! \zeta \partial_\gamma u_\gamma \delta_{\alpha \beta}  ] \!+\! s_{\alpha\beta}  ,
	\end{split}
	\label{density}
\end{equation} 
where $\eta$ and $\zeta$ are the shear and bulk viscosities, $P_{\alpha \beta}$ is the fluid pressure, and $F_\alpha$ is the force density exerted by the polymer on the fluid. 
In our system, the lattice spacing is $1\,\mathrm{nm}$ and time step is $0.0003\,\mathrm{ns}$. 
In order to speed up the diffusive dynamics~\cite{C3SM27410A,Ollila2011-wg}, we set the viscosity and density of the fluid at a value that is 1/10th of that of water at a temperature of $T=350$\,K. Additionally, we kept the kinematic viscosity of the fluid the same as that of water.

In our simulations, monomers and colloid (the pushing sphere) consisted of two types of MD particles: 1) a spherical shell of $N$ nodes to give a well-defined hydrodynamic radius and 2) a central atom, 
Fig.~\ref{box}b) and c). The FENE and LJ potentials are mediated through the central atoms and the spherical shell nodes mediate the hydrodynamic interactions as they interact only with the fluid and give them a well-defined hydrodynamic size. Further details can found in Refs.~\cite{44ref_p,27ref_p, 39ref_p}.  This model has been successfully used to study a wide variety of polymer systems where hydrodynamics plays a key role~\cite{C8SM01445K,doi:10.1021/acs.biomac.3c00473,Ollila2011-wg,C3SM27410A,Ollila2014-xg}

\section{Results}

\subsection{Equilibrium}

\begin{figure}[tb]
		\includegraphics[trim= 350 0 0 0cm,scale=0.37]{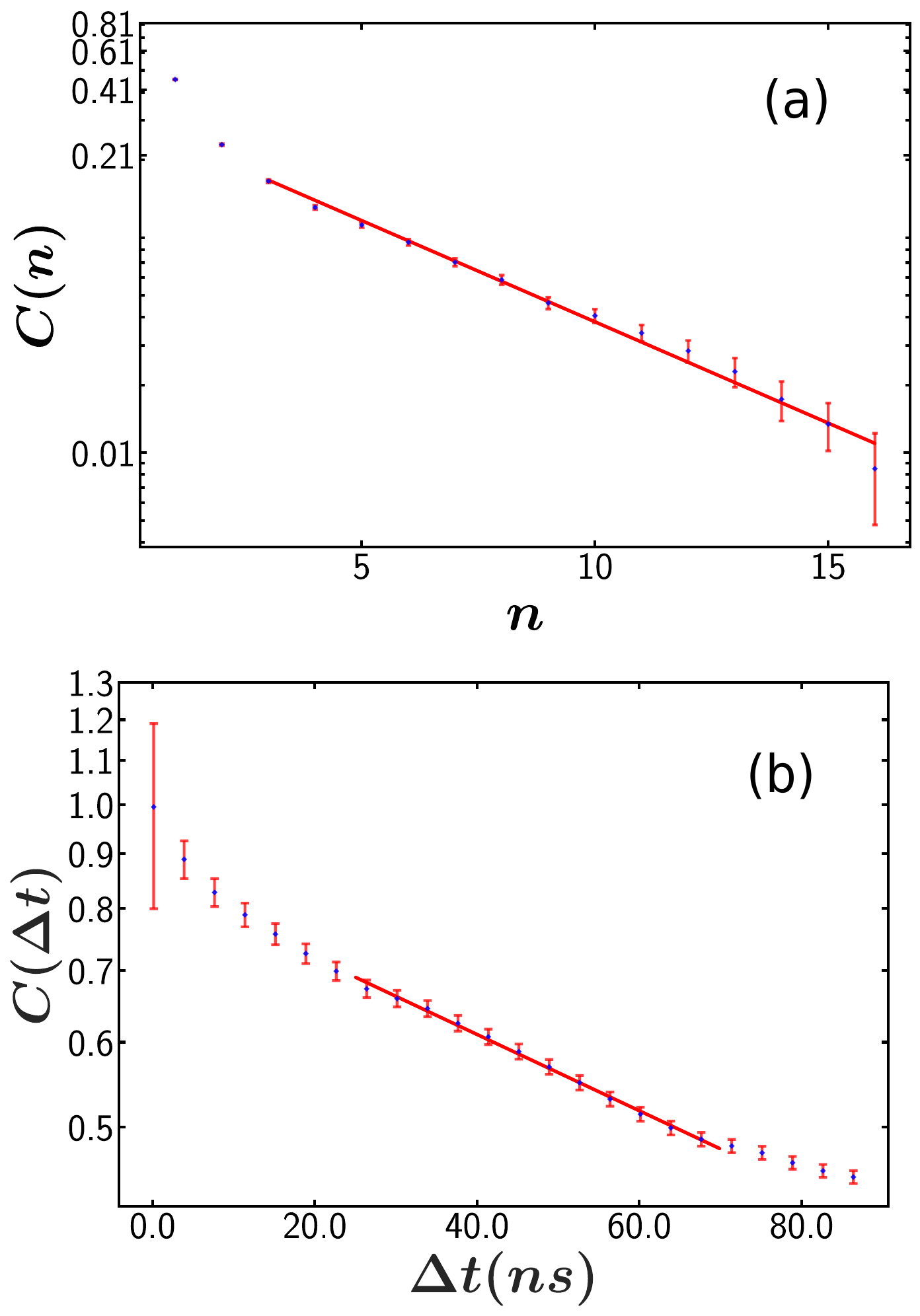}
		\caption{a) The time autocorrelation of the end-to-end distance of the polymer (log scale) as a function of lag time. The red line represents 
  a fit to an exponential decay
  which was used to find the the relaxation time ($\tau$), b) Bond autocorrelation $C(n)$ (log scale) versus the monomer number $n$. The red line represents a fit to an exponential decay 
  which was used to find the chain's persistence length.
		}
		\label{polymer_tau_rep}
\end{figure}

In order to determine whether the confined chain is in the Odijk or the de Gennes regime, the persistence length of the chain in bulk fluid (i.e. not in a channel) was measured from the bond autocorrelation function 
$C(n) = \langle \cos\theta_{i, i+n} \rangle= \langle \vec{v}_i \cdot \vec{v}_{i+n} \rangle$.
$C(n)$ measures 
the average cosine of the angle between the bond vector $\vec{v_i}$ and a bond vector that is $n$ bonds away~\cite{Statistical}.
This function can be approximated as $C(n) \approx \exp\left(-\frac{n l_\mathrm{B}}{P}\right)$, where $l_\mathrm{B}$ is the average bond length.
Figure~\ref{polymer_tau_rep}b) shows $C(n)$ and the corresponding fit. 
The fit yielded
$l_\mathrm{B}= 16.3$\, nm, resulting in a ratio of $D/P$=1.53. Given that $D/P$ is greater than 1, our system falls within the de Gennes regime. 

Although our square nanochannel geometry is similar to that in Bernier \textit{et al.}~\cite{doi:10.1021/acs.macromol.7b02748}, the system is in a distinctly different confinement regime. Bernier \textit{et al.} explored the Odijk regime, where $D/P$ ranges from 0.16 to 0.96. They maintained a fixed channel diameter, and polymer contour length of $D/L=0.015$. In our work, $D/L$ is larger with a value of 0.098. 
The polymer studied by Hayase \textit{et al.}~\cite{PhysRevE.95.052502}  is also in the Odijk regime, characterized by $D/P$ and $D/L$ of approximately 0.01, where the persistence length is of the same order of magnitude as the contour length. Moreover, in their study, the polymer was confined in a channel with a circular cross-section. The channel in our system has a square cross-section, and it has been shown that the de Gennes scaling for square channels is a special case of the more general scaling relations derived for rectangular channels by Werner and Mehlig~\cite{Werner2015-fz}. 

Werner and Mehlig observed that the chain's average extension remains relatively constant regardless of the channel's aspect ratio, as long as one of the dimensions is considerably larger than the polymer's Kuhn length. Meanwhile, the variance in extension is significantly influenced by each channel dimension separately. Therefore, analyzing it in terms of the effective channel size, $\sqrt {D_\mathrm{W} D_\mathrm{H}}$, where $D_W$ is the width and $D_\mathrm{H}$ is the height of the rectangular channel, would be incorrect. 

Polymer confinement within a rectangular channel, but in the Odijk regime, has also been discussed by Muralidhar \textit{et al.}~\cite{Muralidhar2016-tg}. In that work, the condition was specified as $D \leq 2P $ with the channel height being the smaller dimension. The Odijk regime was also studied by Yang \textit{et al.}~\cite{Yang2007-kz} for a polymer fluctuating in a narrow cylindrical channel of diameter $D$ as well as in a channel with its rectangular cross-section in the regime $D \! \ll \! P \! \ll L$ which is different from our system.

\begin{figure*}[tb]
	\includegraphics[trim= 770 0 0 0cm,scale=0.35]{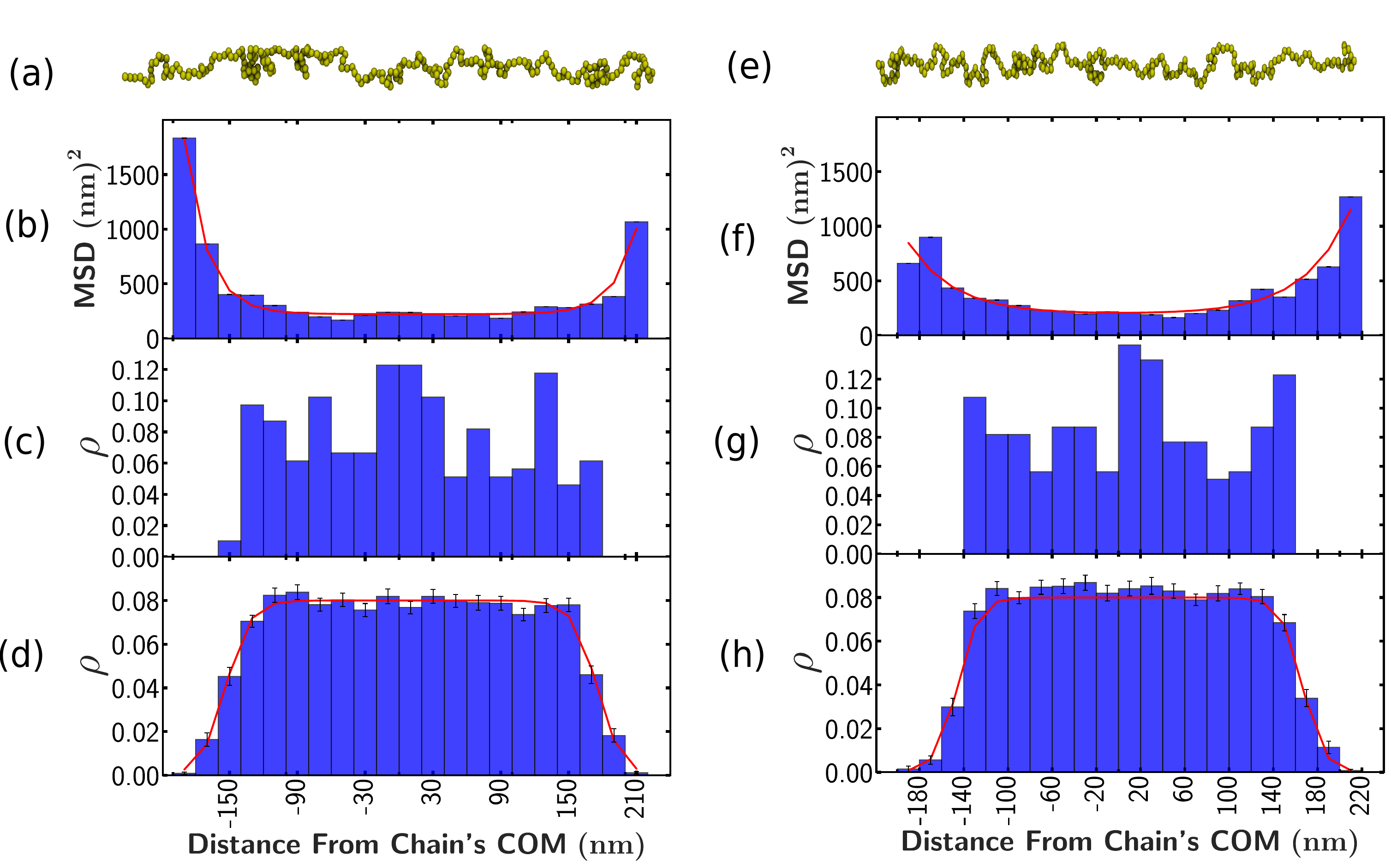}
	\caption{Polymer conformation, average MSD, and the density distribution. The first column (a, b, c, d):
 confined polymer in channel with purely repulsive interactions. The second column (e, f, g, h):  
attractive interactions applied between the monomers. The MSD distributions in (b) and (f) were fit to 
Eq.~\ref{eq:cosh}. The fitted curves are shown in red. 
The instantaneous density distributions in (c) and (g) reveal the distinct blobs along the chain. However, when 
averaging over 4,800 ns, the average density distribution appears flat (d and h).
}
	\label{only_polymer}
\end{figure*}

First, we consider the equilibrium situation with no pushing sphere present in the channel and the polymer with purely repulsive interactions between non-adjacent monomers.  Before data analyses,
each system was equilibrated for about 3,000\,ns.  
The chain's relaxation time was obtained by calculating the 
auto-correlation $C(\Delta t)$ of the end-to-end distance.
Typically,
$C(\Delta t) \propto \exp(-\Delta t/\tau)$, where $\tau$ is the relaxation time of the polymer~\cite{Doi1986-cz}. 
$C(\Delta t)$ is shown in Fig.~\ref{polymer_tau_rep}a)
and the relaxation time was found to be about $ \tau = 121.9\pm 2.8\, \mathrm{ns}$.

Figure~\ref{only_polymer} shows typical configurations of the chain (a),
the average mean squared displacement (MSD) of individual monomers during a time interval equal to the chain's relaxation time (b), and instant (c) and time-averaged density distributions (d); density was measured as number of monomers per distance along $x$.

Blob formation is evident in the instantaneous density distribution depicted in Fig.~\ref{only_polymer}c, which corresponds to the configuration shown in Fig.~\ref{only_polymer}a. In Fig.~\ref{only_polymer}c, 
blobs are manifested as density peaks in the multi-modal distribution. Due to 
thermal fluctuations,
blobs form and dissipate along the chain, and thus the density distribution pattern fluctuates over time. When the distribution was averaged over 4,800\,ns (Fig.~\ref{only_polymer}d), the density fluctuations 
became less noticeable.

The chain also exhibits significant fluctuations in length. 
This results in an increase in monomer-wise MSD 
as a function of distance from the chain's center-of-mass
with the chain ends displaying the largest MSD, Fig.~\ref{only_polymer}b.
The length of the chain at the instant of time that the instantaneous density histogram in (c) is sampled is directly related to the range of the histogram.  This histogram has a narrower range (fewer bins along the horizontal axis) than the histograms in (b) and (d) as a single sample of the polymer is unlikely to have the longest extension sampled in the other histograms averaged over very many configurations. 

The density distribution in Fig.~\ref{only_polymer}d was averaged over 4,800\,ns.  Long time averages do not qualitatively change the density histogram for the equilibrium case.  However, in the next sections where the chain will be pushed along the channel, the time averaged density is no longer flat.  In that case, due to the long wavelength fluctuations in the chains length, 
long averaging times can smear out transition regions of the density profile along the chain.  On the other hand, the instantaneous density is still dominated by the large fluctuations due to the blobs.  Hence, we chose an intermediate averaging time (over $300 \,\mathrm{ns}$, roughly 2.5 times the chains relaxation time) for later density histograms as this is long enough to smooth out fluctuations due to the blobs without smearing out transitions in regions of different density along the chain. The equilibrium density ($\rho_0$) of 0.062 was determined by averaging the density histogram over the intermediate period of 300\,ns and over its entire range.

 We have identified the relevant time scales for the system from the chains relaxation time.  We now look at the possible different length scales.  One length scale can be obtained from the average of the equilibrium MSD (MSD$_0$) in Figs.~\ref{only_polymer}b~and~f.  This gives  MSD$_0$ of $404 \,\mathrm{nm}^2$ for b and $394 \,\mathrm{nm}^2$, or length scales ($\sqrt{MSD_0}$) of $20\,\mathrm{nm}$ and $19.9\,\mathrm{nm}$, respectively.  The MSD plots also show distinct variation at their ends and it is worthwhile to compare the length scale over which this variation occurs to the length scale from $\sqrt{MSD_0}$.  Phenomenologically, interfacial regions often display exponential-like variations.  Motivated by this, we fit the MSD histogram shown in Fig.~\ref{only_polymer}b to the function 
 \begin{equation}
  A + B \cosh((x-x_0)/\lambda),
\label{eq:cosh}
 \end{equation}  
 which we can see fits reasonably well.  Interestingly, the length scale obtained from fitting is $\lambda = 20$\, nm, identical to the value obtained from $\sqrt{MSD_0}$.  
 We can also obtain a length scale from the density histogram transition from $\rho_0$ to zero.  Again, assuming an exponential-like variation, here we fit Fig~\ref{only_polymer}d to the function
 \begin{equation}
  A(\tanh((x-x_0)/\lambda_\rho)-\tanh((x-x_1)/\lambda_\rho),
\label{eq:tanh}
 \end{equation}  
 which we can see fits reasonably well in the figure.  The length scale obtained from this fit was $\lambda_\rho = 21$\,nm.  Once again, this length scale is nearly identical to $\sqrt{MSD_0}$.  This strongly suggests that other than the persistence length there is just a single length scale in the system.  We also note that this length scale is comparable to the channel width (25\,nm) (or more precisely, the channel width minus the monomer-wall interaction length $\sigma_{m-wall}$, which would give 21\,nm).
  
The peaks in the instantaneous density profile Fig.~\ref{only_polymer}c), 
attributed to the blobs in Fig.~\ref{only_polymer}a, typically have a similar width (i.e. 1-2 $\lambda$). According to de Gennes blob model, a polymer of $N$ monomers can be described as a series of $K$ non-penetrating spheres (blobs) with diameter $D$, each containing $N/K$ monomers. Each blob can be individually described by the Flory theory~\cite{Microfluidics}.  However, 
the narrow channel restricts the chain's shape,
and thus it is natural that the channel width 
defines the blob size and the corresponding length scales seen in all the plots.

\begin{figure*}[tb]
	\begin{center}
		\includegraphics[trim= 1500 0 0 0cm,scale=0.141]{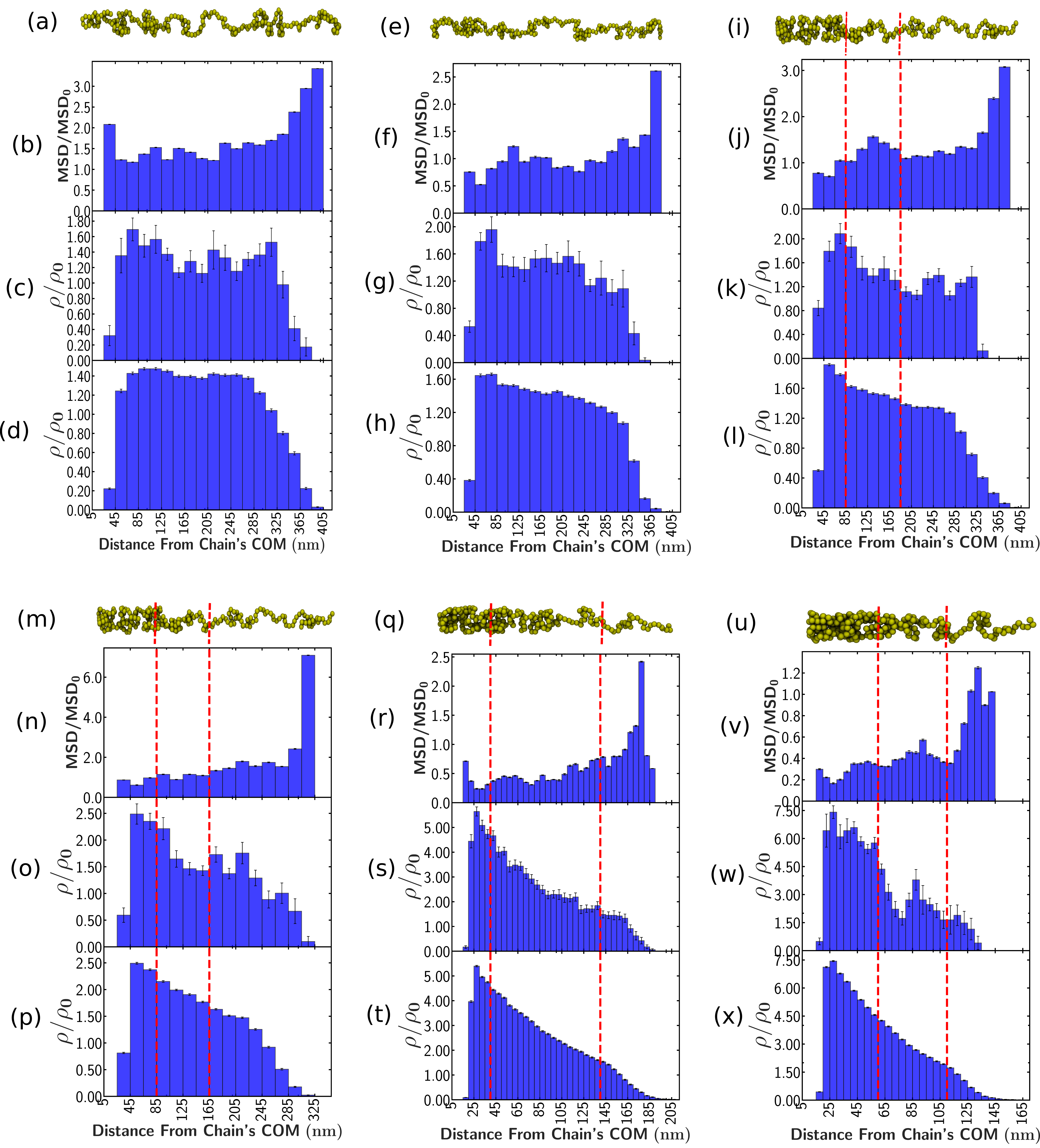}
		\caption{Polymer conformation, average scaled MSD, and density distribution for sphere's velocity of (a, b, c, d) 0.30$v_0$, (e, f, g, h) 1.21$v_0$ , (i,j,k,l) 1.82$v_0$, (m, n, o, p) 3.03$v_0$, (q, r, s, t) 12.12$v_0$, and (u,v,w,x) 24.24$v_0$ for repulsive monomer-monomer interaction. The density histograms in c, g, k, o, s, and w are averaged over a shorter time, but the distributions in d, h, l, p, t, and x are averaged over a longer time. The polymer to the left of the leftmost dashed red line is unambiguously in a higher density, low MSD state whereas the polymer to the right of the rightmost dashed red line is unambigously in a lower density, high MSD state.  Between the dashed lines, there is a crossover where different realizations of the polymer in this region could be in either state.}
		\label{repulsive}
	\end{center}
\end{figure*}

We also performed the same analysis for the case of a weak attractive potential between the monomers. A
snapshot is shown in Fig.~\ref{only_polymer}e.
The instantaneous density distribution in Fig.~\ref{only_polymer}g (corresponding to Fig.~\ref{only_polymer}e)
again exhibits a multi-modal pattern, indicating the existence of denser blobs. Similar to the repulsive case, 
averaging over time smears out the fluctuations, 
Fig.~\ref{only_polymer}h. An equilibrium density of $\rho_0 =0.068$ was determined by averaging the density histogram over 
300\,ns

The attractive chain's relaxation time was determined to be about 106.3\,ns, and the MSD distribution was obtained with lag time equal to the relaxation time. Compared 
to the system with repulsive interactions, the relaxation time for the attractive system is slightly shorter 
and the system has a slightly higher equilibrium density. We again obtain the average of the equilibrium MSD (MSD$_0$) in Fig.~\ref{only_polymer}f which gives  MSD$_0 = 394 \,\mathrm{nm}^2$, or a length scales ($\sqrt{MSD_0}$) of $19.9\,\mathrm{nm}$. Figure~\ref{only_polymer}f shows the MSD distribution and a fit to Eq.~\ref{eq:cosh} and Fig.~\ref{only_polymer}h shows the density distribtuion and a fit to Eq.~\ref{eq:tanh}.  As before, all the length scales obtained from the fits are nearly identical to $\sqrt{MSD_0}$, and the width of the channel.

\subsection{Pushed Polymer}
\subsubsection{Purely repulsive interactions}
\label{repulsive}

In order to examine the chain's deformation under compression, a spherical colloid was used to push the polymer with different velocities from the back (cf. Fig.~\ref{box}a).
Khorshid {\it et al.}~\cite{18ref_p} saw what appeared to be different compressed states as a function of the speed of the pushing bead in an experimental realization of a very similar system.  Part of our goal here is to identify the nature of these compressed states.  One possibility is that these states are thermodynamic phases like a coil-globule, or a liquid-gas, system where compression can, for example, turn a gas into a liquid.  Systems of simple particles with purely repulsive interactions exhibit only one fluid phase while adding attractive interactions can lead to both liquid and gas phases.  We expect that if these compressed phases are thermodynamic phases of the system, then they should be strongly affected by the addition of attractive interactions.  We will investigate this in the next subsection.  

\begin{figure*}[tb]
	\includegraphics[trim= 1355 0 0 0cm,scale=0.19]{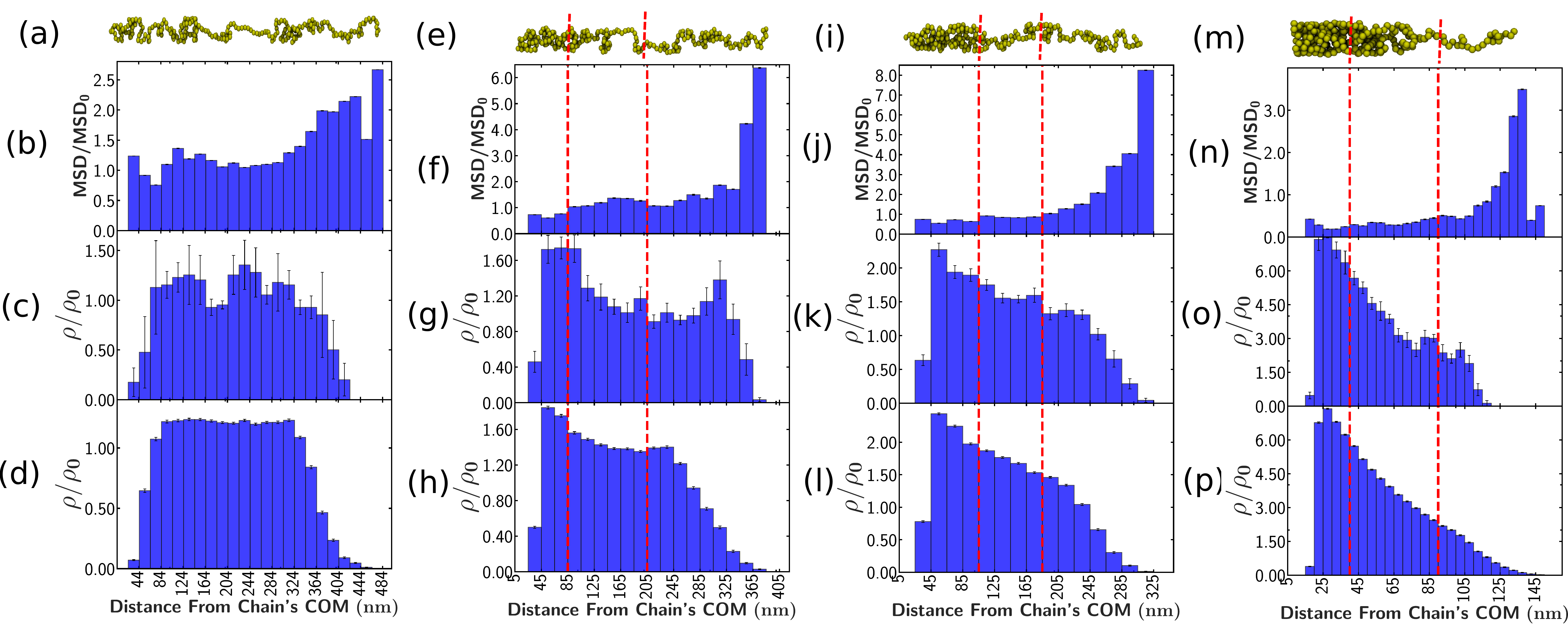}
	\caption{Polymer conformation, average scaled MSD, and the density distributions for sphere's velocity of (a, b, c, d) 0.27, (e, f, g, h) 1.61, (i, j, k, l) 2.67, and (m, n, o, p) 21.39 for attractive monomer-monomer interaction. The density histograms in c,g, and k are averaged over an intermediate period (300 ns) time, but the distributions in d, h, l, and p are averaged over a longer time.  The dashed red lines have a similar meaning as in Fig.~\ref{repulsive}.}
	\label{attractive}
\end{figure*}

Another possibility is that the compressed states are entirely dynamic phases of the system, such as the "jamming" seen in granular flow in a pipe\cite{Brand2011}.  A jammed state may be disordered, but is typically characterized by a higher density (i.e. compressed) and low relative motions of the particles.  We now investigate the effect of gradually increasing the velocity of the spherical colloid on the state of compression of the polymer.  A natural velocity scale for the problem is $v_{0} = \sqrt{(\text{MSD}_{0})}/\tau$  which is equal to 0.165\,nm/ns and 0.187\,nm/ns for the repulsive and attractive cases, correspondingly.  

For the repulsive case, Figs.~\ref{repulsive}a,e,i,m,q,~and~u show snapshots of a typical polymer  configuration when moving the sphere with gradually increasing velocities from 0.30$v_0$ to 24.24$v_0$.  At low velocities the polymer configuration does not show significant qualitative differences from the equilibrium configuration seen in Fig.~\ref{only_polymer}a.  However, at velocites of 1.82$v_0$ and above, there is a distinct compacted region of the polymer at the back (where it is being pushed by the sphere).  We now look at these configurations in more detail. 

First, we consider moving the sphere with velocities of 0.30$v_0$ (Figs.~\ref{repulsive}a-d) and 1.21$v_0$ (Figs.~\ref{repulsive}e-h).
The density distributions in Figs.~\ref{repulsive}c,g 
and Figs.~\ref{repulsive}d,h
resemble the equilibrium case except that the polymer is slightly compressed as indicated by the larger than equilibrium density values in the central region. At the back end (close to the sphere), the density distribution drops to zero faster than in the equilibrium case whereas the leading edge of the polymer far from the sphere shows a slightly slower drop.  
There is a much more dramatic change in the MSD profiles (cf. Figs.~\ref{repulsive}b,f vs. Fig.~\ref{only_polymer}b). The large peaks at the chain ends in Figs.~\ref{only_polymer}b,f are strongly suppressed are strongly suppressed due to long wavelength fluctuations.
The regions of higher density corresponding to a more compressed portion of the chain have slower monomer movement, as indicated by lower MSD.  However, the MSD values are mostly above the equilibrium value of
$404.36\,\mathrm{nm}^2$, suggesting enhanced fluctuations, over and above the thermal Brownian-like motion present in equilibrium.  This is likely a function of the fact that the polymers experience a non-uniform flow due to the motion.
With no fluctuations, and in the absence of the polymer, this would create a Poiseuille-like flow profile ahead of the sphere. However, the polymer mostly fills the channel and its interaction with the flow (often referred to as "backflow") results in a more plug-like flow with large hydrodynamic fluctuations coupled with large fluctuations in polymer density.  
Similar to the equilibrium case, blobs are formed and then dissipated, but they last longer at the back end of the chain. This is consistent with the lower MSD and higher density in this region of the chain.  At the front (farthest way from the sphere), the density is low and the corresponding MSD is high.

At a velocity of 1.82$v_0$ (Figs.~\ref{repulsive}i-l), a sign of state crossover is observed. The blobs against the sphere at the back end of the chain become slightly larger and denser with higher density and lower MSD values.
For a velocity of 3.03 $v_0$ (Figs.~\ref{repulsive}m-p), the chain's conformation becomes quite different with a stronger indication of a crossover. 
Two separate parts are observed in the distributions and the chain's conformation. There is a small region at the back end of the chain with high density, while the other larger part is at the front end of the chain with slightly lower density.
The results indicate that a crossover occurs due to distinct regions in the density and MSD distributions.  
At the crossover speed both dense and non-dense states appear to coexist. The dense state, the part of the chain that has the highest density, has the lowest MSD values. This indicates that the monomers do not move around much and only have small distance fluctuations. The blobs are long-lived and no longer dissipate in this region. In contrast, the non-dense state, the part of the chain with the lowest density, has the highest MSD values, that is, 
the monomers have more mobility, they can move, and beads still form and dissipate in this region. As these two separate parts can be seen in the distributions, we can conclude the two states coexist around the crossover speed.

When the sphere's velocity increases further to 12.12$v_0$ (Figs.~\ref{repulsive}q-t) and 24.24$v_0$ (Figs.~\ref{repulsive}u-x),
the presence of two distinct states becomes more distinguishable in the MSD and density distributions. 
The density distributions reveal the presence of a dense state at the back end of the polymer 
and non-dense state at the front end of the chain similar to the lower speeds above
the crossover speed.

\subsubsection{Weakly attractive interactions}
\label{attractive}

Above, we considered the case of purely repulsive monomers.
A similar crossover is observed in the case of weakly attractive monomers, but at different sphere velocity. Like in the repulsive case, a few higher density "beads" are formed along the chain at the velocity of 0.27$v_0$ (Figs.~\ref{attractive}a-d). 
The results indicate that generation and dissipation of  beads occur over time.  Signs of two-state existence start to appear at v=1.61$v_0$ (Figs.~\ref{attractive}e-h).
The denser and larger blobs are formed at the back end of the chain. There is, however, a slightly less dense part observed at the front end of the chain. 
The crossover becomes more distinguishable for velocities greater than 2.67$v_0$ (Figs.~\ref{attractive}j-l). 
As the sphere's speed increases to a very high value of 21.39$v_0$ (Figs.~\ref{attractive}m-p),  the distinction between the two states becomes more apparent especially in the density histogram. 

Overall, there is not a lot of difference between the behaviour of the system with purely repulsive interactions between non-adjacent monomers, and the system with attractive interactions.  As mentioned above, this adds weight to the argument that the compressed states we see are entirely dynamic phases of the system.

\subsubsection{Radial distribution functions}

\begin{figure}[tb]
	\advance\leftskip-9cm
	\includegraphics[width=0.5\textwidth]{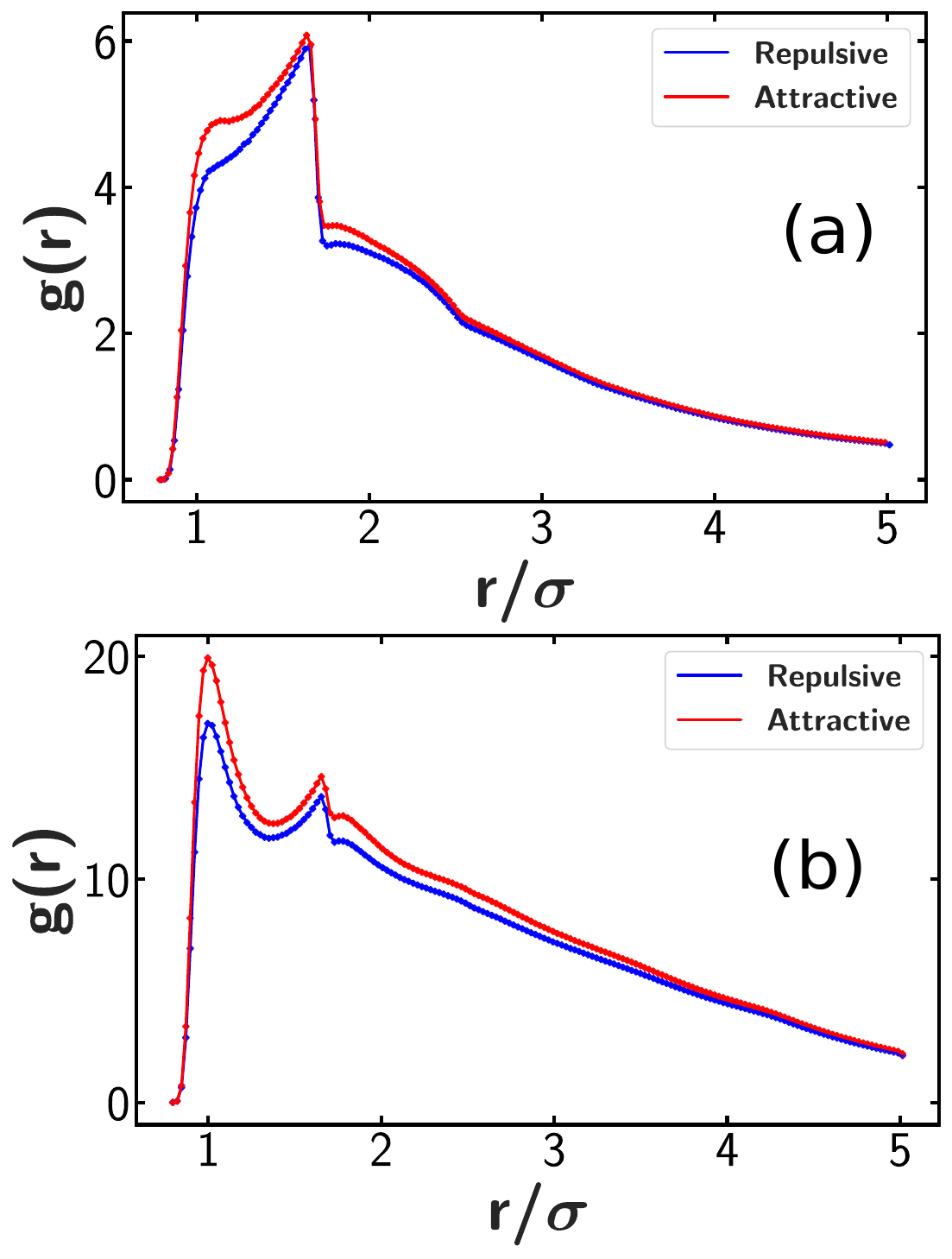}
	\caption{Radial distribution function
 versus the pairwise separation distance between the monomers ($r$), which is scaled by $\sigma$, for velocity a) below the crossover point(0.61$v_0$ for repulsive and 0.53$v_0$ for attractive and b) above the crossover point (60.61$v_0$ for repulsive and 53.48$v_0$ for attractive). 
	}
	\label{g_r}
\end{figure}

To understand the arrangement of monomers in both attractive and repulsive scenarios, we examined the radial distribution function (RDF) denoted as $g(r)$. 
Figure~\ref{g_r} illustrates the relationship between $g(r)$ and the pairwise distance $r$ for the velocities of 0.61$v_0$ and 60.61$v_0$ for the repulsive case, and 0.53$v_0$ and 53.48$v_0$ for the attractive case. The velocities are above and below the crossover, respectively. Our results demonstrate that the attractive case exhibits a higher peak in $g(r)$ for both velocities compared to the repulsive case. This illustrates 
a slightly more organized arrangement in the case of the attractive potential.

There is a clear difference in $g(r)$ at the velocities well above and below the transition in Fig.~\ref{g_r}.  At high velocity, the first peak (around $r/\sigma=1$) is higher than the second peak whereas the second peak is higher at low velocities. Due to the coexistence of the two states (so that any $g(r)$ measurement averages over the behaviour of both states), it is hard to tell if this enhanced caging happens precisely at the transition. However, it does indicate that the monomers are more caged by their nearest neighbors in the compressed state seen as the velocity is increased. 

\subsubsection{Characterization}

\begin{figure*}[tb]
	\advance\leftskip-17cm
	\includegraphics[width=0.95\textwidth]{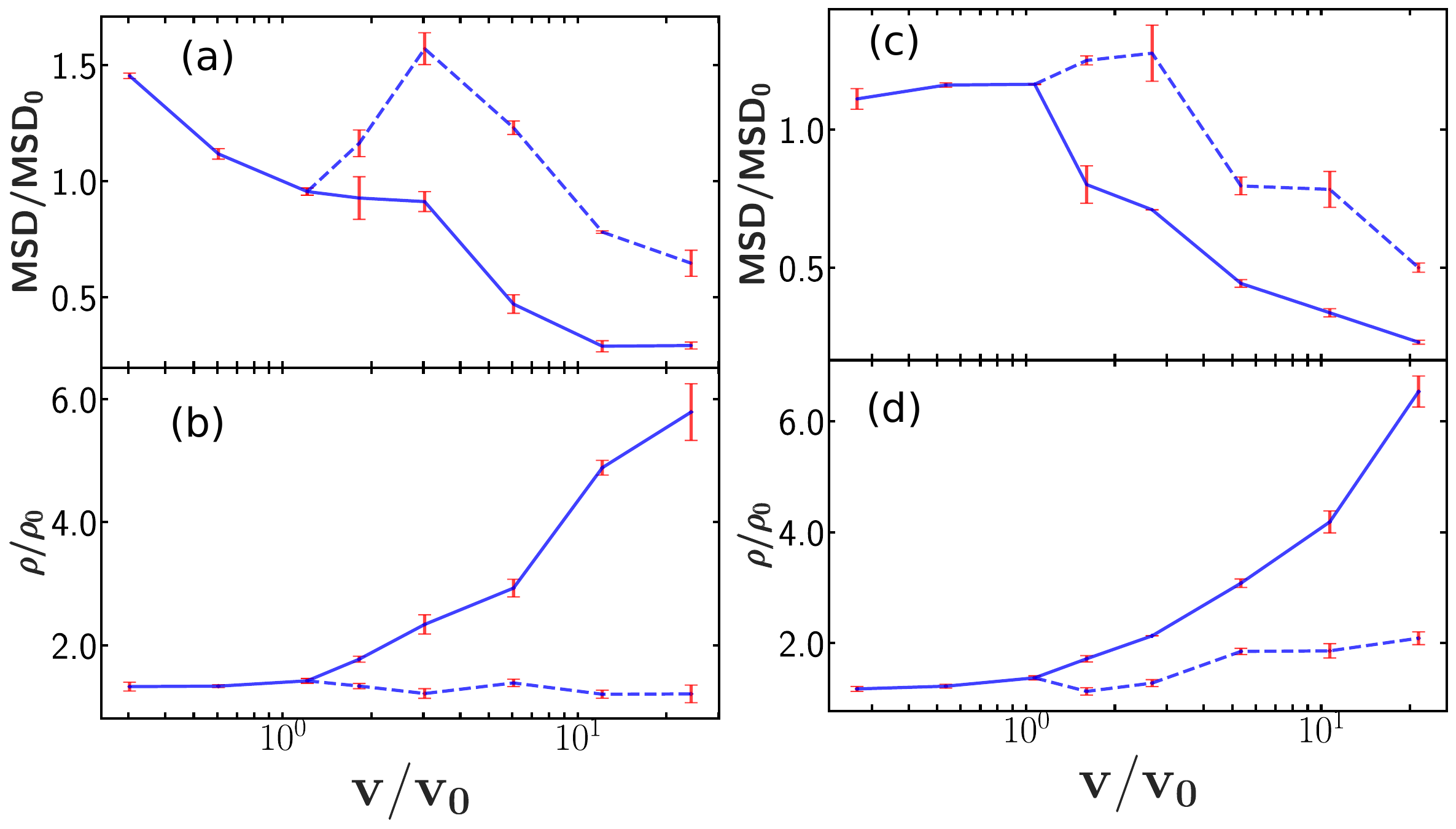}
	\caption{MSD and volume per monomer as a function of sphere's velocity for a) and b) repulsive, and c) and d) for attractive potential. For $v < v^*$, a single state exists, showing up as one branch in volume per monomer and MSD. Beyond $v^*$, two branches emerge: a lower branch for the dense state (low MSD, low volume per monomer) and an upper branch for the non-dense state (high MSD, high volume per monomer). $v^*$ is 4.42$v_0$ for repulsive and 3.36$v_0$ for attractive monomer interaction.
	}
	\label{msd_volume}
\end{figure*}

To characterize the presence of the two states, we calculated the average density and the average MSD for regions unambiguously in each state and plotted these values as a function of $v/v_{0}$ for both the repulsive and the attractive cases in Fig.~\ref{msd_volume}.  When the velocity is lower than the crossover, 
we considered the density distribution and excluded the bins with high density fluctuations, which typically occur at both ends of the chain.  The same process was applied to compute the average MSD values. 

For velocity above the crossover point, we partitioned the two distinguishable parts in density and MSD distributions corresponding to the dense and teh non-dense state.
This is shown in Fig.~\ref{repulsive}. We found the mean of the MSD values and the density within specified bins for each dense and non-dense regions separately. 
Bins characterized by high fluctuations in both MSD and density were excluded when computing the mean values and the results are scaled by the equilibrium value. 

When the sphere's velocity is less than chain's equilibrium velocity, the diffusive motion of the monomers is faster than the convective motion of the sphere,  and for these speeds we see no sign of chain compaction. However, when the sphere's velocity overcomes the monomer's diffusive velocity $v_0$, the diffusive motion 
cannot keep up the convective motion of the sphere leading to compact chain conformation. This crossover becomes obvious at $v^*$, which is 1.82$v_0$ for repulsive and 1.61$v_0$ for attractive monomer interaction.

For $ v < v^*$, there is only one state existing along the chain which is consistent with only one branch for density and MSD. For $ v > v^*$, two branches appear indicating the two-state coexistence. The dense state corresponds to the lower branch with low MSD, and the non-dense state is related to the upper branch with higher MSD and volume per monomer. A sign of the crossover can also be detected in the chain's dynamics. For both repulsive and attractive cases, blobs were generated and they moved in the direction reverse with respect to the chain's center of mass. They  dissipated over time for $ v < v^*$, i.e. they were dragged back towards the sphere due to their size). This process occured repeatedly during the run time. However, the chain's dynamics becomes different for $ v > v^*$. At higher speeds, the beads travelled in reverse direction of the chain but then became dumped against the sphere, and were not able to dissipate in that region. Therefore, this process leads to the preference for the dense state to exist next to the sphere in the two-state crossover. 
In both the dense and the low density states, MSD generally decreases as a function of the sphere's velocity.  The exception is just above the crossover where there is a modest increase in the MSD for the low density state.

\begin{figure}[tb]
	\advance\leftskip-10cm
	\includegraphics[width=0.5\textwidth]{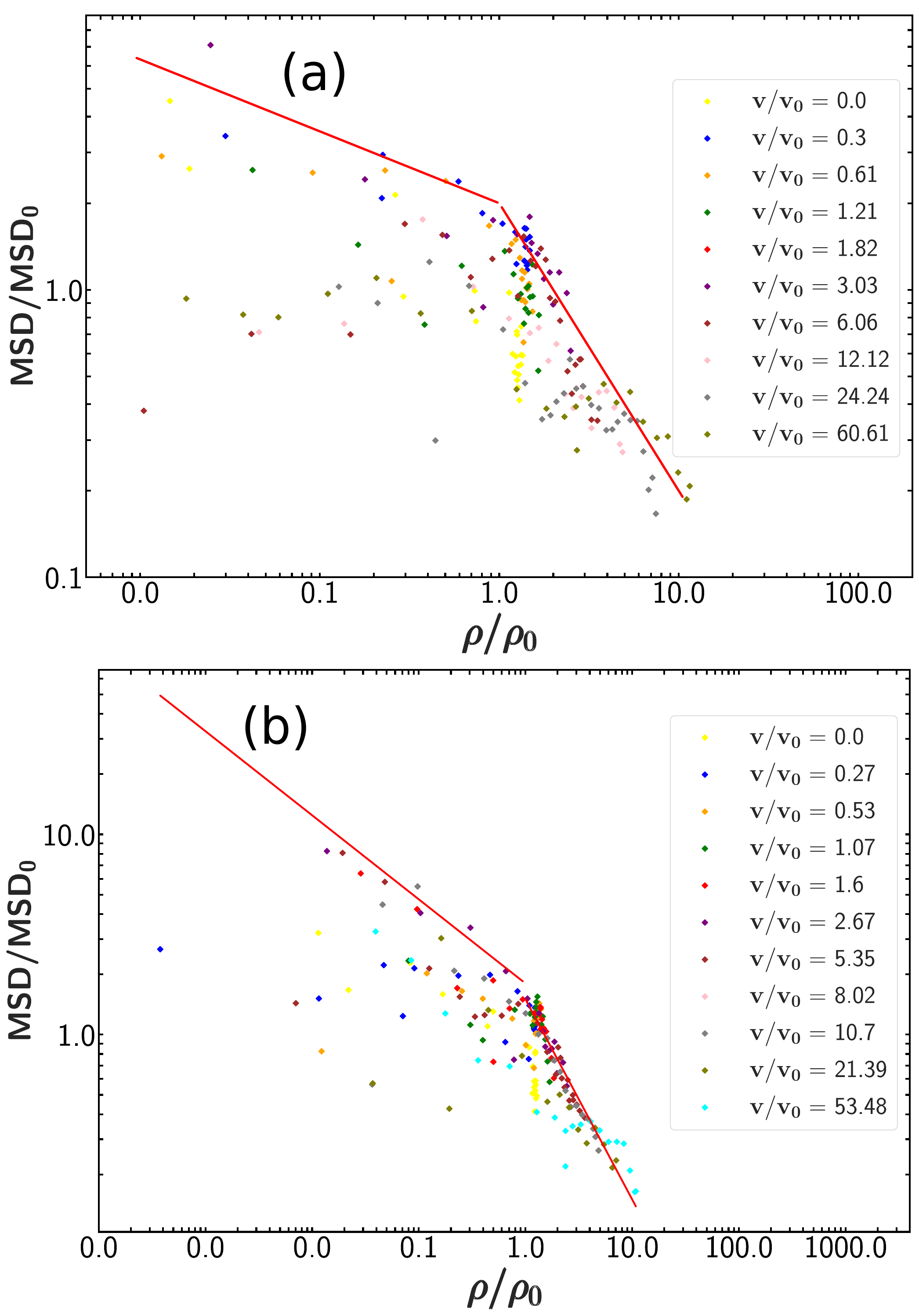}
	\caption{MSD versus density (log scales) for a) repulsive and b) attractive potential. The plot reveals that there is an inverse correlation between MSD and density. The estimated relationship for MSD/{MSD}\textsubscript{0} is expressed as $c/(\rho/\rho_0)^a$, depicted by the red line. The analysis indicates that for the repulsive scenario, $c=2.0$ and $a=0.25$, while for the attractive scenario, $c=1.8$ and $a=0.42$ apply to densities below the equilibrium density ($\rho < \rho_0$). Conversely, for densities surpassing the equilibrium density ($\rho > \rho_0$), the repulsive case is characterized by $c=2.0$ and $a=1.0$, and the attractive case is represented by $c=1.5$ and $a=1.0$.
	}
	\label{msd_rho}
\end{figure}

We also constructed a plot to depict the relationship between the MSD and density for all bins, Fig.~\ref{msd_rho}. Despite the presence of noise in the plot, a general trend emerged indicating an inverse correlation between the MSD and density. To quantitatively characterize this relationship, we estimated that the ratio MSD/{MSD}\textsubscript{0} goes as $c/(\rho/\rho_0)^a$ (or, at least the upper envelope of the data follows this trend). For densities below the equilibrium density ($\rho < \rho_0$), we obtained the exponents $a=0.25$ and $a=0.42$ for purely repulsive and weakly attractive monomers, respectively. Conversely, for densities exceeding the equilibrium density ($\rho > \rho_0$), the fitting yielded $a=1.0$ for both the repulsive and attractive cases.  Below the transition, the results are clearly non-universal, but for the dense state the exponent appears independent of the details of the inter-monomer potential and  is suggestive of a jammed state. 

\subsection{Polymer Folding}

\begin{figure*}
	\advance\leftskip-18cms
	\includegraphics[width=0.9\textwidth]{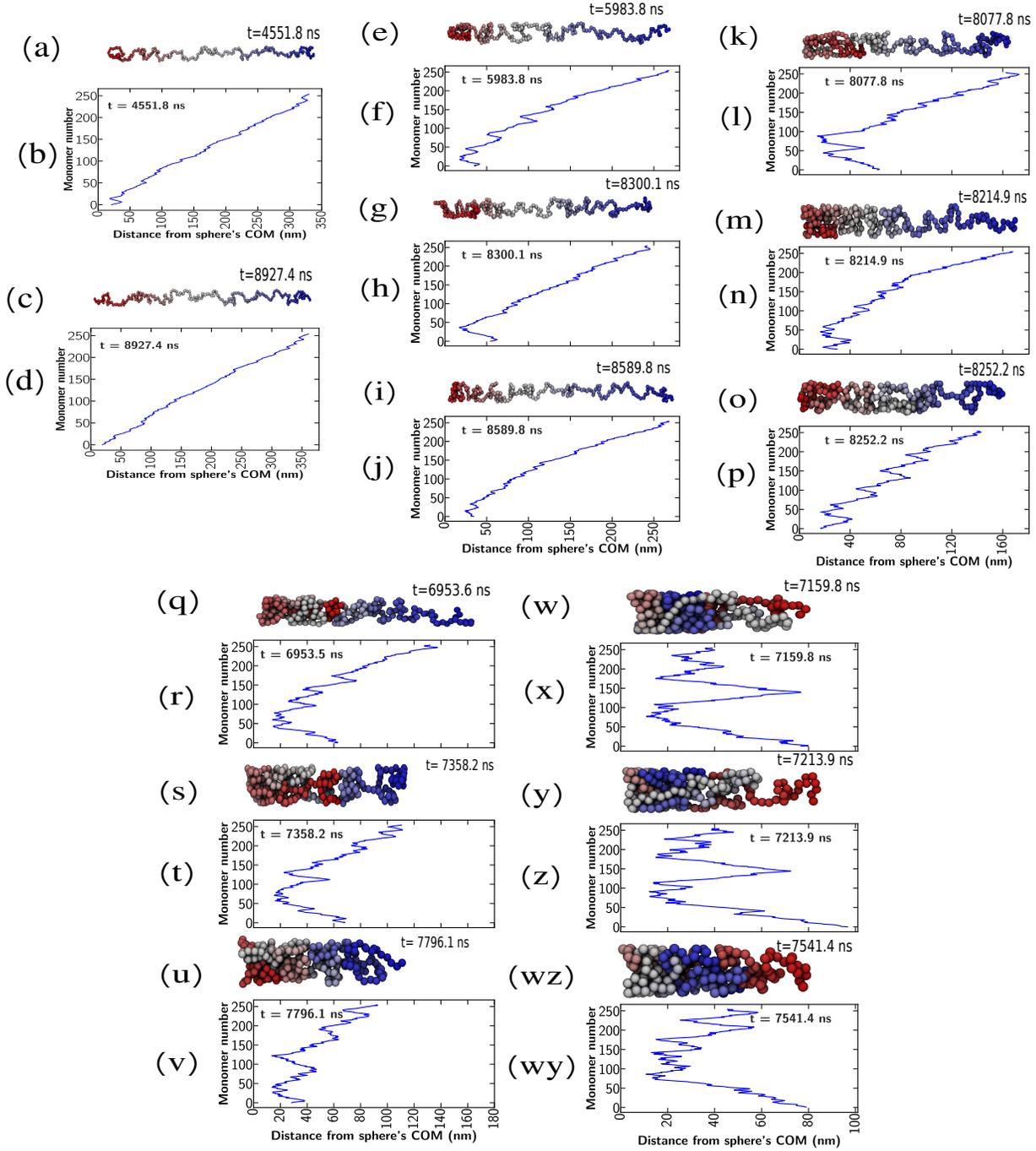}
	\caption{The chain structures and the monomer number versus the distance from sphere's COM for sphere's velocities of a-b) 0.30$v_0$, c-d) 1.21$v_0$, e-j) 3.03$v_0$, k-p) 12.12, q-v) 24.24$v_0$ and w-wy) 60.61$v_0$ at a few different times during the simulation. The monomers are colored from blue to red based on the monomer index (1-255) along the chain. 
	}
	\label{fig:structures}
\end{figure*}

To further explore how the polymer configures itself, we examined folding of the chain as it translates along the channel with the sphere moving at the back end of the chain. We assign monomer numbers based on their sequence in the chain starting with the end closest to the sphere.
The monomer number versus the distance from the sphere's center of mass is plotted  in Fig.~\ref{fig:structures}b for the case where the sphere moves with velocity 0.30$v_0$ 
at $t=$4551.8\,ns. The chain structure is also shown in Fig.~\ref{fig:structures}a where it is colored from red to blue according to the monomer number.  A kink in the plot 
represents a fold in the chain's structure. When the chain translates along the channel, only a few blobs are nucleated and move in the reverse direction with respect to the center of mass of the chain and then become dissolved. In order to measure the lifetimes of the blobs, a sequence of instantaneous density histograms were generated, and the time over which the density bumps existed was used as an estimate for the lifetime. The estimated lifetime for the blobs at the back end of the chain was determined to be about 1.2\,ns and no folding occurred. 
The blobs correspond to the regions where the polymer segments have small fluctuating bends or random zigzags. Based on our observations we, determined that one reason for the  higher density close at the back edge of the polymer is due to  the higher number (compared to equilibrium) of irregular small bends in the polymer segments close to the sphere is higher.

At $v=$1.21$v_0$, the recurring process of generation and dissipation of local blobs was observed. This is similar to the results at lower velocity. The main difference is that the number of zigzags associated with the local blobs increases close to the sphere,
Figs.~\ref{fig:structures}c~and~d,  and that is why the density at these segments is higher.  At the opposite end of the polymer, segments far from the sphere are able to move more freely. The lifetime of the local blobs, 1.8\,ns, is also slightly longer than at the lower velocity.

Figures~\ref{fig:structures}e-j show 
the monomer number as a function of distance from the sphere's center of mass for the velocity of 3.03$v_0$, which is greater than $v^*$, at random times during the simulation time after equilibration.
The figure shows 
that small fluctuating folds are generated along the chain from the middle of the chain toward the back of the chain at $t=5983.8$\,ns. 
These small local folds are associated with blobs starting to get stuck in this region and not dissipating. As a result, the density of this region is higher due to the presence of folds/blobs. At $t=8300.1$\,ns, only one single small fold/loop nucleates at the back end of the chain against the sphere indicating that most of the blobs that arrive in this region are not allowed to dissipate, and that there is only one region that contains most of the blobs making the density of the corresponding region higher compared to the other times. 
This is consistent with the crossover to a dense state that was explained through the MSD and density histograms. In the $t=8589.8$\,ns snapshot, this small fold close to the sphere breaks out into small segments and then gets nucleated again. This process happens repeatedly as the sphere keeps pushing the chain along the channel. The motion of the sphere creates a hydrodynamic flow along the chain and therefore the polymer experiences a hydrodynamic force. The competition between this hydrodynamic force and the excluded volume interaction between the monomers determines when folding occurs. When the compression hydrodynamic force overcomes the bending force, folding with blobs that do not dissipate takes place. If the hydrodynamic force is slightly greater than the bending force folding occurs, but it can again be broken into small zigzags. It is worthwhile mentioning that the folds 
are irregular and the resulting structure is not organized. 
As discussed earlier, Bernier \textit{et al.}~\cite{doi:10.1021/acs.macromol.7b02748} observed the dynamics and the folding structures to be disordered as the edge of the Odijk regime is reached ($D/P > 0.5$).
Connecting polymer segments between folds fluctuate greatly in length even though folding is still evident and irregular sequence of folding events takes place with folds formed at random positions and times. Bernier \textit{et al.} also speculated that $D/P > 1$ corresponds to different compression physics leading to packing of disordered blobs. In our system, 
$P=16.3$\,nm and therefore the ratio of the channel width ($D=25$\,nm) to the persistence length of the chain is greater than one which indicates that we are in de Gennes regime. Here, irregular folding occurs along with formation of the blobs along the chain.

At the higher velocity of 12.12$v_0$ at $t=8077.8$\,ns, more than one fold containing zigzags are nucleated close to the sphere. The corresponding plots can be found in Figs.~\ref{fig:structures}k-p. During the simulation, double folding is observed which results in loop generation. The larger zigzags and folds along the chain close to the sphere correspond to the dense state. On the other hand, the region with small zigzags at the front edge of the chain corresponds to the non-dense region. 
In the $t=8214.9$\,ns snapshot, we see the large double folds are broken into smaller folds and these folds gets larger at $t=8252.2$\,ns, and the whole process happens again repeatedly. 

As the velocity increases to 24.24$v_0$, at $t=6953.6$\,ns, more folds with slightly longer strands are created from the middle of chain towards its back edge, 
Figs.~\ref{fig:structures}q~and~r. At $t=7358.2$\,ns, two folds containing small zigzags are nucleated against the sphere. These folds consist
of longer strands with small bends along them. 
In addition, loops were observed when more than one fold was created. Figures~\ref{fig:structures}s~and~t show the corresponding chain structures as well as a plot of monomer beads along the channel's axis.
At $t=7796.1$\,ns, the chains are slightly shorter and the folds are also created toward the front end of the chain, Figs.~\ref{fig:structures}u~and~v. The compression flow increases as a function of speed which results in a dense folded irregular structure at the edge close to the sphere. The non-dense region at the front end of the chain also contains the folds that make the density and MSD slightly higher and lower correspondingly at this region.

At $v=60.61 v_0$ and $t= 7159.8$\,ns, a larger fold is present at the middle section of the chain. The red monomers which were initially close to the sphere move toward the front of the chain which makes the strands of the fold longer,  Figs.~\ref{fig:structures}w~and~x. At $t=7213.9$\,ns, Figs.~\ref{fig:structures}y~and~z indicate that the central fold at the middle of the chain becomes smaller while the red monomers move farther from the sphere. Later at $t=7541.4$\,ns the larger central fold is broken into smaller folds (Figs.~\ref{fig:structures}wz~and~wy). The results show that at very high speeds the front end of the chain tends to be compacted too:
the dense region containing more folds grows to the front of the chain which makes the lower-density region shorter and slightly denser.  It is also clear that the longest folds have very long lifetimes (the longest folds in Fig.~\ref{fig:structures}x are still present in Fig.~\ref{fig:structures}wy).  These long folds essentially result in four or five parallel segments of the chain being confined in the same region of the channel.  Such parallel segments can be packed quite efficiently leading to the high monomer caging seen in the $g(r)$ plot, Fig.~\ref{g_r}b. It is also this higher effective level of confinement (from multiple chains being confined to the same region of the channel) that effectively reduces the long time mobility of the monomers.

\begin{figure}[tb]
	\advance\leftskip-8.75cms
	\includegraphics[width=1.\columnwidth]{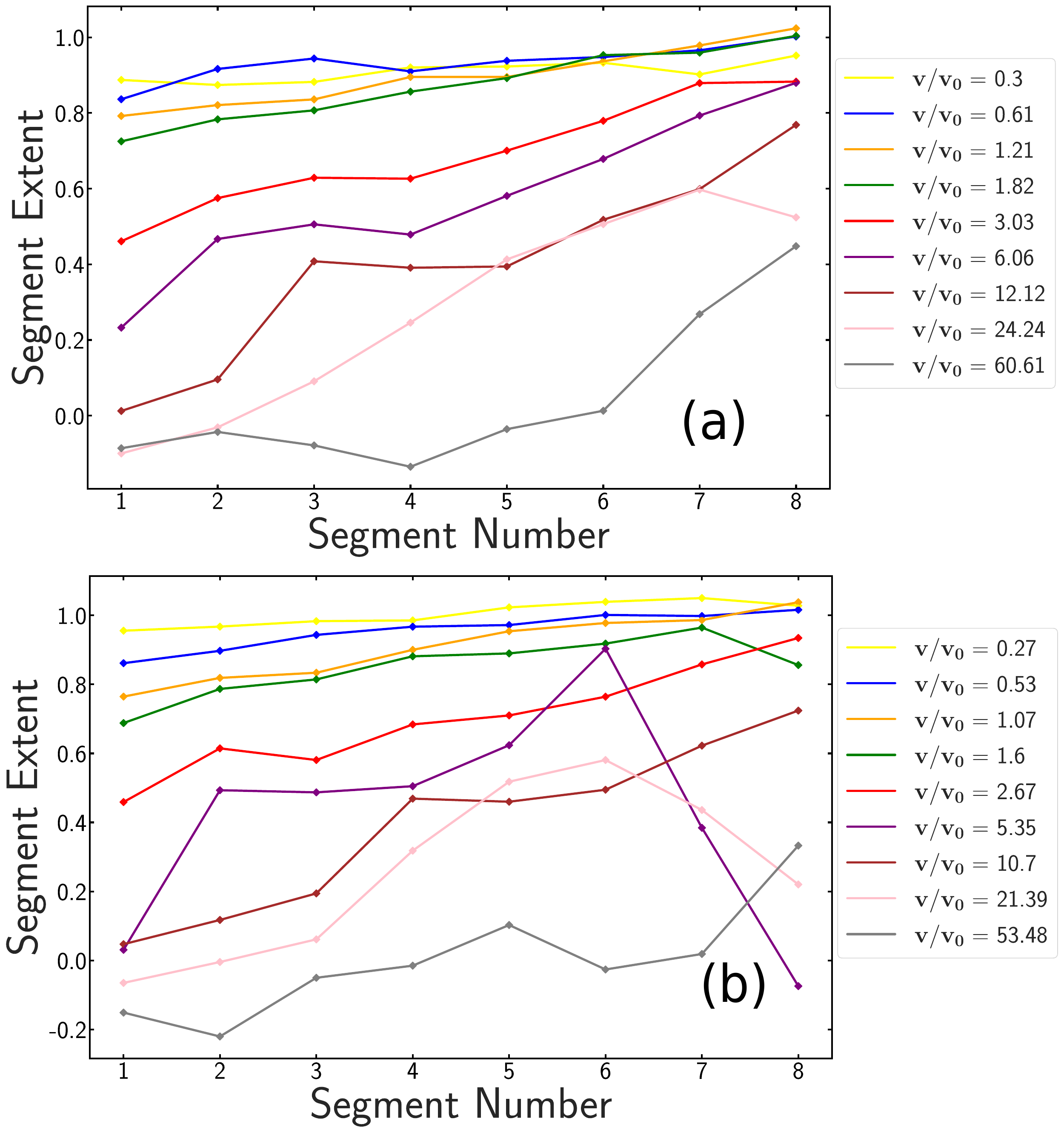}
	\caption{The polymer is divided into 8 segments and the extent along the $x$-direction (distance from sphere's centre-of-mass) is plotted versus the segment number at different speeds of the sphere for (a) purely repulsive monomer interactions and (b) attractive monomer interactions.  Segment 1 is closest to the sphere and segment 8 is farthest.
	}
	\label{fig:seg_exten_num}
\end{figure}

\begin{figure}[tb]
	\advance\leftskip-8.75cms
	\includegraphics[width=1.\columnwidth]{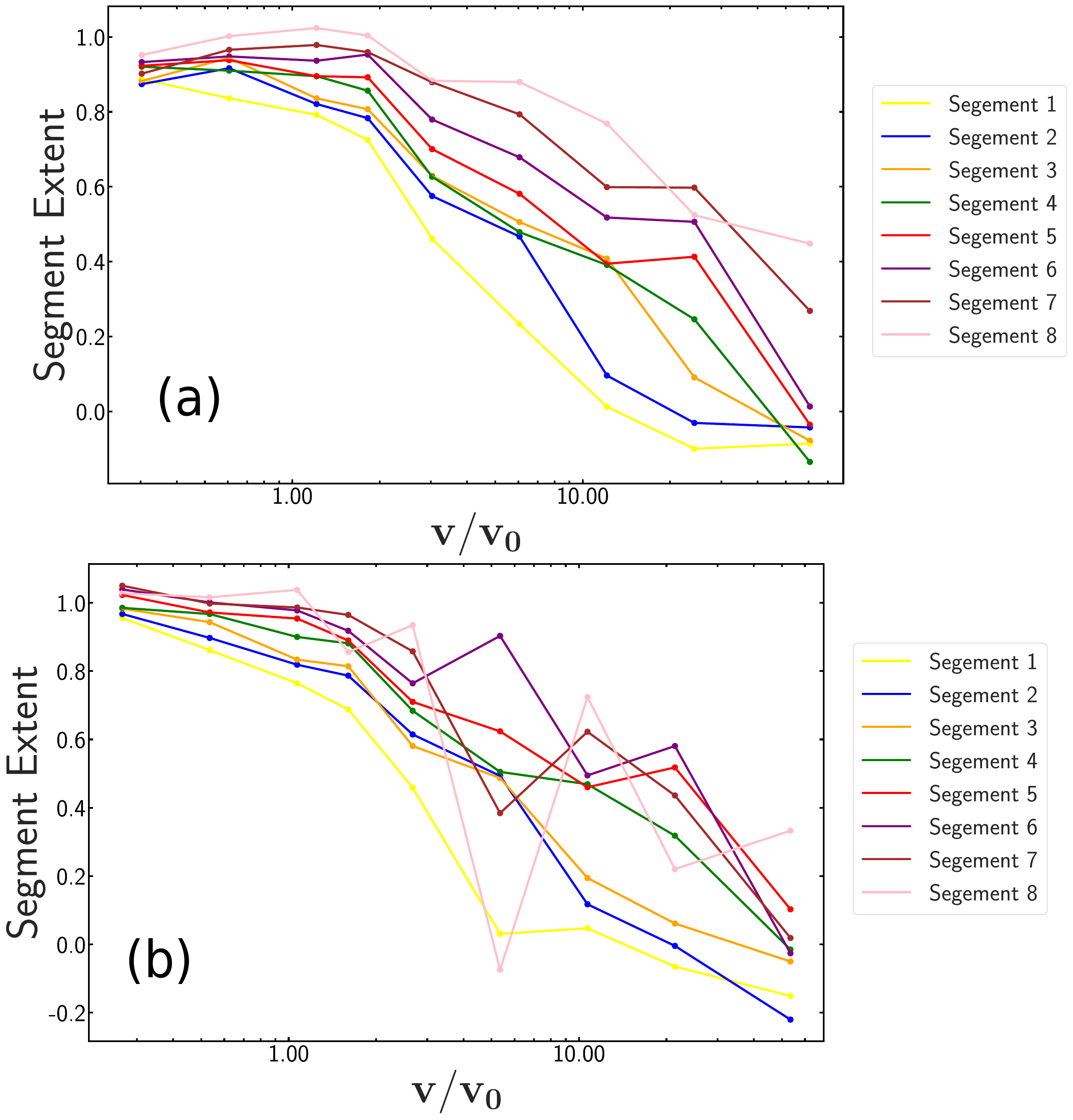}
	\caption{The extent of each polymer segment as a function of sphere speed for (a) purely repulsive monomer interactions and (b) attractive monomer interactions.  The segment extent is normalized by the average segment extent in equilibrium.
	}
	\label{fig:seg_exten_V}
\end{figure}

To quantify the effects of polymer folding, we divide the polymer into eight segments, excluding 16 monomers from each end.  The segments are numbered from 1, for the segment closest to the sphere, to 8 farthest from the sphere.  We plot each segment's extent along the $x$-direction (difference in distance from the sphere's COM to the monomer for the last and the first monomer in the segment) as a function of segment number in Fig.~\ref{fig:seg_exten_num}.  Note that when there are folds, the segment could have either positive (extending away from the sphere) or negative extent (folded back so the segment is extending towards the sphere).  The extent is averaged over time, and for the high velocity data averaged over up to 10 independent realizations of the system.  As the folded states can have both positive extents at a given instant of time, or given realization of the system, the average over folded states such as those seen at high velocity in Fig.~\ref{fig:structures} will average to zero.  In Fig.~\ref{fig:seg_exten_num} we see that at low velocities all segments have a similar extent, close to that of a segment in an equilibrium chain.  As the sphere velocity increases, we see the segment extent drop towards zero for segments closest to the sphere that are compacted and folded.  This region of low extent gradually extents to more and more segments as the velocity increases.  Figure~\ref{fig:seg_exten_V} shows the segment extent versus the sphere velocity for each segment.  These curves have a sigmoidal shape and the crossover from equilibrium extent to near zero extent happens first in segments close to the sphere and last in segments furthest from the sphere.
From this analysis we clearly see that the folding transition enables the compaction to the dense jammed state of the polymer at high velocities.

\section{Knots, or rather, Their Absence}

In the experiments of Khorshid \textit{et al.}~\cite{Amin2018} a sliding gasket in a nanochannel
was used to investigate the transient crossover in confined polymers. The conformations, however,  could not be directly followed due to lack of available tools. The bright spots in the experiments are referred to as regions of high density, and, consequently, interpreted as the location of knots. 

We examined compressed polymers to detect possible knotted regions. Mathematically speaking, the topological knotted state of a circular chain is unique and well-defined. This is attributed to the property that the knotted state remains unchanged under distortions or alterations in the chain's geometry, as long as the connectivity of the
chain is maintained. To apply the idea of knottedness to open chains, it is necessary to close the open chain in a ring with an auxiliary arc~\cite{doi:10.1021/ma048779a, doi:10.1143/PTPS.191.192}. As a result, the knotted state of the ring can be assigned to the open chain. 
This can be implemented by considering a minimally interfering closure scheme~\cite{doi:10.1143/PTPS.191.192}. In this scheme, two distinct ways of connecting the chain ends are compared: 1) via their direct connection, or 2) the closest points of the convex hull of the chain portion. The end-to-end distance is then compared with the summed distances of the termini from the convex hull. If the former is smaller, direct bridging is used, otherwise the closure is through the convex hull. We used the KymoKnot software~\cite{Tubiana} to analyze knots in our systems. In KymoKnot, the minimally interfering scheme closure is used to identify knots along the polymer.

The analysis shows that no knots were formed along the chain when it was being pushed by the sphere, and only the folding probability increased as a function of the sphere's velocity. This result is different from the findings of Roth\"orl \textit{et al.}~\cite{Rothorl2022-lm}. 
Roth\"orl \textit{et al.} used Langevin simulations to study knot formation in a model of confined dsDNA being pushed by a nano-dozer in a nanochannel whose width is much smaller than the contour length of the polymer. At low stiffness, when the polymer is compacted with increasing velocity, the probability of overall knotting increased too. The speed of the piston was used to tune the occurrences of knots. At slow piston velocities towards the equilibrium state, the decrease of knotting was accompanied by a trend towards a weak localization of trefoil knots. They mentioned that the configurations with a higher total knotting probability are formed at high stiffness because the high density induces the formation of multiple or more complex knots, and the amount of trefoil knots is reduced. At high persistence length, compactification leads to backfolding which, on the other hand, creates loops -- a prerequisite for knots. Here in our work, we observed backfolding only at high sphere velocity, and dense
regions did not necessarily appear to be correlated with knot formation. 

Our results also differ from those of Ruskova \textit{et al.}~\cite{Ruskova2023-bk}. They conducted coarse-grained MD simulations employing a Langevin thermostat on DNA polymers being pushed through infinite open chiral and achiral channels. They explored the polymer's behaviour in the channels by examining metrics such monomer distributions, and changes in the topological state of the polymer. Focusing on cylindrical open channels, their analysis indicated that the probability of knotting increases with the velocity of pushing. Furthermore, there was a shift toward the occurrence of more complex knots with an increasing pushing velocity, and they mentioned that this observation is consistent with the previous investigations of Roth\"orl \textit{et al.}~\cite{Rothorl2022-lm} 
Both Roth\"orl \textit{et al.} and Ruskova \textit{et al.} employ Langevin simulations with an implicit solvent without incorporating 
hydrodynamic interactions. In addition, compaction was driven by a drag force proportional to the velocity of the pusher. However, our approach involves LB simulations with an explicit fluid and with hydrodynamic interactions.  In particular, the fluid is pushed ahead of the sphere and the polymer induces back-flow in the fluid resulting in a much more complex and nuanced compaction force than the simple drag seen in the Langevin simulations. It is worth noting that hydrodynamic interactions also play a crucial role in such systems~\cite{C3SM27410A, Dorfman2014-vb}; 
in Ref.~\cite{C3SM27410A},
statics and dynamics of a flexible polymer confined between parallel plates were investigated considering both the presence and absence of hydrodynamic interactions. The planar diffusion coefficient for the center of mass (CM) of the chain, $D_\parallel$ decreased with increasing confinement, aligning with experimental observations. However, it is worth noting that Langevin dynamics simulations do not accurately capture the reduction in $D_\parallel$ with increasing confinement, making them less suitable for studying dynamic phenomena in confined environments. 

To compare our results with those in Refs.~\cite{Rothorl2022-lm, Ruskova2023-bk}, we conducted Langevin simulations for the polymer with solely repulsive interactions. The polymer was pushed in the channel by a moving sphere similar to the system we had in the LB simulations. In LD simulations, where $\Delta t = 0.007$\,ns, the damping parameter in the frictional drag equation ($F_r = -\frac{m}{\text{damp}}v$) was set to $20 \tau_{\text{lj}}$\,ns and T=350\,K. Here, $m$ represents the mass of the monomer, which was set equal to $1.3044\,\text{ag}$. The term $\tau_{\text{lj}} = \sigma \sqrt{\frac{2m}{\epsilon}}$, where $\sigma$ and $\epsilon$ are the same LJ parameters employed in the LB simulations. It was observed that no knots were formed along the chain at any sphere velocities.
This result demonstrates that the presence or absence of hydrodynamic interactions does not affect knot formation. The reason for the difference in our results from those in Ref.~\cite{Ruskova2023-bk}  might be attributed to the fact they employ a harmonic bond instead of a FENE bond, which allows for easy bond crossings, leading to the generation of knots. Additionally, in Ref.~\cite{Rothorl2022-lm}, the channel width
was 240\,nm and the persistence length
50\,nm, resulting in a ratio of $D/P$=4.8. This ratio is significantly larger than the one in our work (1.53).

\section{Conclusions}

In this paper, multi-state crossovers of a confined polymer in a nanochannel were studied. The chain was immersed in a fluid, and a sphere, which is present at the back end of the chain, was used to generate compression. The velocity of the sphere was varied and chain's behaviour was examined through density and MSD distributions.  

We observed variations of density with velocity and distance from the pusher that match those reported in the experiments of Khorshid {\it et al.}~\cite{18ref_p}.  At low velocities of the pusher, the distribution of density was observed to be similar to that of an equilibrium chain.  Slightly above the characteristic velocity $v_0$ of monomer diffusion,  multi-modal density and MSD distributions emerged.  The part of the chain close to the sphere was in a dense state with low monomer displacements over time, compared to the density and $\sqrt{MSD_0}$ of an equilibrium chain. The front end of the chain was observed to be in a low-density state with higher monomer displacements (comparable or even higher than the equilibrium $\sqrt{MSD_0}$). 
The crossover occurred for both repulsive and attractive monomer-monomer interaction.  Other than a {\it slightly} different $v^*$ there we did not observe significant differences between the behaviour of the system with purely repulsive versus attractive monomer-monomer interactions.. 

We also compared the radial distribution function for the compact and low density states.  In the compact state, the monomers were tightly caged by other (non-adjacent along the chain) monomers while the monomers were poorly caged in the equilibrium and low-density states.

The crossover was also examined through the study of folding events. At very low sphere velocity, only small zigzags were created along the chain which corresponds to small blobs created along the chain. However, for velocities greater than the crossover velocity, larger folds were created at the part close to the sphere. These folds led to parallel/anti-parallel segments of the chain going in both directions leading to effectively higher levels of confinement. At very high sphere speeds, larger long-lived folds with longer strands were observed.

The lack of any dependence on the nature of the inter-monomer interactions strongly suggests that the observed crossover is not related to equilibrium states of the polymer chain.  The fact that the crossover occurs at a velocity scale close to the diffusive velocity of monomers in an equilibrium chain suggests that this is a dynamic crossover from a region dominated by diffusion at low velocities to a region dominated by advection driven compaction at high velocities.
Since the jamming crossover is a  a phenomenon that occurs when the density of a disordered assembly of particles is increased, and in which the dynamics slows down dramatically to the point where the system can no longer relax and becomes a rigid state that acts like a solid under compression~\cite{doi:10.1038/nphys580,PhysRevE.84.011303}, the dense region against the sphere at high sphere's velocity appears to be related to the jamming crossover~\cite{jamming_book,PhysRevE.84.011303}.
This is also consistent with the tight caging of monomers seen in the jammed state that is made possible by the long folds present in the chain.  The jammed state we see, while folded, is free of knots suggesting that knotting is not required to explain the results of the experiments of Khorshid {\it et al.}~\cite{18ref_p}.

\begin{acknowledgments}
	MK and CD thank the Natural Science and Engineering Research Council of Canada (NSERC) for financial support.
	MK also thanks the Canada Research Chairs Programs
	of the Natural Sciences and Engineering Research Council of Canada (NSERC) and SC thanks the Ontario Graduate Scholarship (OGS) for financial support. 
	This research has been enabled by the use of computing resources provided by Shared Hierarchical Academic Research Computing Network (SHARCNET) and the Digital Research Alliance of Canada.
\end{acknowledgments}


\bibliography{apssamp}

\end{document}